\documentclass[11pt,a4paper]{article}
\usepackage{jstyle}

\def\a{\alpha}
\def\b{\beta}

\def\d{\delta}

\def\e{\epsilon}

\def\m{\mu}

\def\cK{{\cal K}}
\def\cL{{\cal L}}

\author[]{Euihun JOUNG\quad }
\author[]{Luca LOPEZ\quad }
\author[]{Massimo TARONNA}

\affiliation[]{Scuola Normale Superiore and INFN\\
Piazza dei Cavalieri 7, 56126 Pisa, Italy}

\emailAdd{euihun.joung@sns.it}
\emailAdd{luca.lopez@sns.it}
\emailAdd{massimo.taronna@sns.it}

\title{\center
Solving the Noether procedure for cubic interactions of higher spins in (A)dS}

\abstract{The Noether procedure represents a perturbative scheme to construct all possible consistent interactions starting from a given free theory. In this note we describe how cubic interactions involving higher spins in any constant-curvature background can be systematically derived within this framework.
}

\addnote{Invited contribution to the J.~Phys.~A special volume on ``Higher Spin Theories and AdS/CFT'' edited by Matthias Gaberdiel and Misha Vasiliev.}

\begin{document}

\maketitle

\section{Introduction}

String theory (ST) and Vasiliev's equations (VE) \cite{Vasiliev:1988sa,Vasiliev:2003ev} are
the only known examples of consistent theories
of interacting higher-spin (HS) particles.\footnote{
See \cite{Bekaert:2010hw,Sagnotti:2011qp} for some recent reviews.}
Although their current formulations provide mathematically elegant descriptions
involving infinitely many auxiliary fields,
some important aspects, as the number of derivatives 
 involved in the cubic vertices or the possible (non-)local
 nature of higher-order interactions, are hidden.
 Indeed, the cubic vertices involving fields in the first Regge trajectory
 of the open bosonic string
have been obtained only recently in \cite{Taronna:2010qq,Sagnotti:2010at}.
Leaving aside Chan-Patton factors, their illuminating form is given by
\be
	\left|V_{3}\right>=
	\exp\left\{\tfrac12\sum_{i\neq j} \sqrt{2\a'}\,
	a^{\dagger}_{i}\cdot p_{j}
	+a^{\dagger}_{i}\cdot a^{\dagger}_{j}
	\right\}\,\left| 0 \right>_{123} \qquad
	\left[(a^{\mu})^{\dagger}\equiv \a_{-1}^{\mu}\right],
\ee
where \mt{i,j=1,2,3} label the Fock spaces associated to the interacting particles.
From the latter expression, the \mt{s_{1}\!-\!s_{2}\!-\!s_{3}} interactions,  as well as the corresponding coupling constants, can be easily extracted via a Taylor expansion of the exponential function.
Besides reflecting the world-sheet (Gaussian integral) origin of ST,
the latter fulfills all requirements dictated by 
the compatibility with both the string spectrum 
and the corresponding global symmetries.
Hence, many key properties  
can be deduced by investigating the consistent cubic vertices. Concerning VE, one might ask 
what is the form of the cubic vertices and 
what can one learn from them. 
Given the analogy to ST, we expect to understand
how the global HS symmetry constrains 
the massless \mt{s_{1}\!-\!s_{2}\!-\!s_{3}} interactions  making 
the entire spectrum of VE as a single immense multiplet.
This question has been partly addressed in 
\cite{Sezgin:2002ru,Sezgin:2003pt,Boulanger:2008tg},
where it has been shown that, starting from VE, the extraction of the cubic vertices
requires infinitely many field redefinitions making the analysis very involved.

On the other hand, moving from a top-down to a bottom-up viewpoint, 
one may ask oneself which HS cubic interactions 
lead to fully nonlinear HS theories and 
whether ST and VE are the only solutions or not.
This question (called Gupta or Fronsdal program)
can be tackled solving the Noether procedure  for HS fields. The latter is a perturbative  scheme (order by order in the number of fields) whose aim is to classify all consistent interactions starting from a given free theory.
The first step of such procedure is to find out the most general couplings of 
three massive or massless HS particles 
in an arbitrary constant-curvature background. 
For the case of symmetric HS fields, this problem has been addressed in \cite{Joung:2011ww,Joung:2012rv}, where, making use of the ambient-space formalism, 
all possible 
 \mt{s_{1}\!-\!s_{2}\!-\!s_{3}} interactions were provided.\footnote{
 See \cite{Fradkin:1986qy,Vasilev:2011xf} for a frame-like approach to the problem of 
 massless interactions, and \cite{Buchbinder:2006eq,Zinoviev:2008ck,Boulanger:2008tg,Zinoviev:2010cr,Fotopoulos:2010nj,Bekaert:2010hk,Polyakov:2012qj} for other works on HS cubic interactions in (A)dS.}

Notice that the aforementioned program is closely related to the
classification of all consistent CFTs in arbitrary dimensions.
More precisely, in the context of AdS/CFT, the Noether procedure seems to share many analogies with the problem of classifying all possible consistent OPEs of HS operators $\mathcal O_{i}(x)$\,. In turn,
this is tantamount to enumerating all possible tensor structures 
of three-point functions
\mt{\left< \mathcal O_{i}(x_{1})\,\mathcal O_{j}(x_{2})\,\mathcal O_{k}(x_{3})\right>},
from which all conformal blocks can be computed.\footnote{
See \cite{Giombi:2011rz,Costa:2011mg,Costa:2011dw,SimmonsDuffin:2012uy,
Stanev:2012nq,Zhiboedov:2012bm} for the correlation functions of three conserved currents.}
In this respect,
it would be interesting to understand the dictionary between 
the Noether procedure requirements on the bulk side
and the conformal symmetry on the boundary side.
This perspective can possibly clarify the role of Lagrangian locality, usually assumed in the bulk, or of possible alternatives, and may provide a new look into the AdS/CFT correspondence itself. 
Moreover, it is worth mentioning that the CFT results (3-pt functions) can be obtained \emph{a priori} 
from the AdS results (cubic vertices) 
attaching the Boundary-to-Bulk propagators to the vertices.

In the present paper we present the construction
of HS cubic  interactions in (A)dS along the lines of \cite{Joung:2011ww,Joung:2012rv}. We shall show
that this problem is equivalent to finding polynomial solutions of a 
rather simple set of linear PDEs.
Each solution is in one-to-one correspondence to a consistent cubic interaction.
Let us stress that, since the solution space is linear, an arbitrary linear combination of 
these cubic vertices is also consistent, leaving 
their relative coupling constants unfixed (at this order).
The latter are constrained within the Noether procedure
 either by compatibility with the global symmetries of the free theory 
and/or by consistency of quartic interactions.
Let us stress once again the connection to the conformal bootstrap program which may entail key (still unclear) requirements dictated by the Noether procedure. 
 
The organization of the paper is the following: in Section \ref{sec: Noether} we introduce the Noether procedure which represents the main tool of our construction. In Section \ref{sec: flat} we briefly review how to apply such a scheme to derive cubic interactions in flat space. Section \ref{sec: Amb HS} is devoted to the formulation of the free theories in the ambient space. The ambient-space action at the cubic level is discussed in Section \ref{sec: Amb act}. Finally, in 
Sections \ref{sec: HS cubic} and \ref{sec: sol} we present the solution to the cubic order of the Noether procedure in (A)dS.

\section{Noether procedure}
\label{sec: Noether}

The aim of the Noether procedure is to find all consistent (at least classically)
interacting structures associated to a given set of particles,
order by order in the number of fields.\footnote{
See \cite{Berends:1984rq} for a detailed discussion.}
In the case of massless spin 1 or spin 2 particles ($A_{\mu}$ or $h_{\mu\nu}$), 
this corresponds to identifying the consistent interactions starting from
Maxwell or Fierz-Pauli Lagrangians.
Arbitrary vertices involving $A_{\mu}$ or $h_{\mu\nu}$
would mostly cause a propagation of \emph{unphysical} DOFs, 
which, at the free level, are removed by the linear gauge symmetries:
\mt{\delta^{\sst (0)}A_{\mu}=\partial_{\mu}\alpha} and
\mt{\delta^{\sst (0)}h_{\mu\nu}=\partial_{\mu}\xi_{\nu}+\partial_{\nu}\xi_{\mu}}\,. 
Hence, a key condition for the consistency of the interacting theories is the existence of gauge symmetries which are nonlinear deformations of the linear ones.

Let us consider an arbitrary set of gauge fields $\varphi^{a}$ 
(where $a$ labels different fields) with free action $S^{\sst (2)}$ 
and linear gauge symmetries \mt{\delta^{\sst (0)}\varphi^{a}}\,.
The problem is to find the corresponding nonlinear action $S$ together with the non-linear gauge symmetries
$\delta\varphi^{a}$\,. For this purpose, one can consider the following perturbative expansions:
\ba
	&&S=S^{\sst (2)}+S^{\sst (3)}+S^{\sst (4)}+\cdots\,,
	\label{S}\\
	&&\delta\,\varphi^{a}=\delta^{\sst (0)}\varphi^{a}+\delta^{\sst (1)}\varphi^{a}+
	\delta^{\sst (2)}\varphi^{a}+\cdots\,,
	\label{ddelta}
\ea
where the superscript $(n)$ denotes the number of fields involved.
Taking the variation of the action \eqref{S} under the gauge transformation \eqref{ddelta},
one ends up with a system of gauge invariance conditions:
\ba
	&& \delta^{\sst (0)}S^{\sst (2)}=0\,,
	\label{N1} \\
	&& \delta^{\sst (0)} S^{\sst (3)}+\delta^{\sst (1)}S^{\sst (2)}=0\,,
	\label{N2} \\
	&& \delta^{\sst (0)} S^{\sst (4)}+\delta^{\sst (1)}S^{\sst (3)}+
	\delta^{\sst (2)}S^{\sst (2)}=0\,,
	\label{N3} \\
	&& \hspace{40pt} \cdots \nonumber
\ea
The first equation implies the linear gauge invariance of the free theory.
The second equation is a condition for both the cubic interactions $S^{\sst (3)}$
and the first-order gauge deformations $\delta^{\sst (1)}\varphi^{a}$\,, and so on.
Since the second term in \eqref{N2} is proportional to the
free EOM, condition \eqref{N2} implies
\be\label{N2-}
	\delta^{\sst (0)}S^{\sst (3)}\approx 0\,,
\ee
where $\approx$ means equivalence up to the free EOM. Solving 
this equation one can identify all cubic interactions consistent with
the linear gauge symmetries. In the case of massless spin 1, one finds
two independent interaction terms which schematically read
\be\label{s1 cubic}
	S^{\sst (3)}=
	\lambda^{\sst 1\!-\!1\!-\!1}_{1}\,{\textstyle\int}\,A\,A\,F\ +\ 
	\lambda^{\sst 1\!-\!1\!-\!1}_{0}\,{\textstyle\int}\,F\,F\,F\,.
\ee
The first is the one-derivative YM vertex while the second 
is the three-derivative Born-Infeld one.
In the massless spin 2 case, there are three independent interactions:
\be\label{s2 cubic}
	S^{\sst (3)}=
	\lambda^{\sst 2\!-\!2\!-\!2}_{2}\,{\textstyle\int}\,h\,\partial h\,\partial h\ +\ 
	\lambda^{\sst 2\!-\!2\!-\!2}_{1}\,{\textstyle\int}\,R\,R\ +\ 
	\lambda^{\sst 2\!-\!2\!-\!2}_{0}\,{\textstyle\int}\,R\,R\,R\,,
\ee
where the first is the two-derivative gravitational minimal coupling 
while the other two come from the expansions of (Riemann)$^{2}$
and (Riemann)$^{3}$ and involve four and  six derivatives respectively.
As one can see from these lower-spin examples, 
the general solutions to eq.~\eqref{N2-} 
are \mt{s_{1}\!-\!s_{2}\!-\!s_{3}} vertices with different number of derivatives 
associated with the coupling constants $\lambda_{n}^{\sst s_{1}\!-\!s_{2}\!-\!s_{3}}$\,.
It is worth noticing that 
these coupling constants are independent at this level. 

The next step consists in solving \eqref{N2}
for the first-order gauge transformations  $\delta^{\sst (1)}\varphi^{a}$
associated to each solution $S^{\sst (3)}$ found at the previous step.
In the lower-spin cases, only the first vertices in \eqref{s1 cubic} 
and \eqref{s2 cubic} lead to nontrivial deformations:
\be\label{s1,2 tr}
	\delta^{\sst (1)}A=\lambda^{\sst 1\!-\!1\!-\!1}_{1}\,A\,\a\,,
	\qquad
	\delta^{\sst (1)}h=\lambda^{\sst 2\!-\!2\!-\!2}_{2}\,
	(h\,\partial \xi-\partial h\,\xi)\,,
\ee
of the linear gauge
transformations. The latter
correspond to the standard non-Abelian YM gauge transformations and to diffeomorphisms 
respectively.
Although not deforming the gauge symmetries,
the remaining vertices can be completed 
to the full non-linear order keeping consistency
with the gauge transformations \eqref{s1,2 tr}.
They form the first elements of a 
class of higher-derivative gauge or diffeomorphism invariants,
where the remaining elements appear at higher orders $S^{\sst (n)}$\,. 
General (HS) gauge theories present as well two types of cubic vertices:
the ones deforming the linear gauge symmetries,
and the ones giving rise to possible higher-derivative gauge invariants.
Although the former define the deformed non-Abelian gauge algebra,
the second are also relevant 
since they provide possible (quantum) counter-terms. 
Hence, if no independent non-deforming vertices survive at higher-orders,
then no counter-terms would be available and 
the theory would be UV finite.  
This issue can be initially addressed solving eq.~\eqref{N3}
which involves quartic interactions as well as the spectrum of the theory \cite{Taronna:2011kt,Dempster:2012vw}.

So far we have only considered the case of gauge theories. 
Constructing interactions of massive HS fields also raises similar problems. 
Arbitrary interaction vertices would mostly violate the Fierz conditions
resulting in the propagation of unphysical DOFs.
One way to proceed is to introduce Stueckelberg symmetries into 
the massive theory and perform the Noether procedure
similarly to the massless case.\footnote{
See e.g. \cite{Zinoviev:2008ck} for some explicit constructions.}
%Another equivalent way, that is actually the one we use in the following,
%is to first construct the transverse and traceless parts of the vertices and then consistently complete them.

\section{Flat-space case} 
\label{sec: flat}

In this section we consider flat-space cubic vertices since their construction reveals some of the key ideas also used in the (A)dS case.
See the review \cite{Bekaert:2010hw} 
for an exhaustive list of references on the cubic interactions,
and \cite{Fotopoulos:2010ay,Buchbinder:2012iz,Metsaev:2012uy,Henneaux:2012wg} 
for more recent developments.

In order to deal with arbitrary HS fields, it is useful to introduce  auxiliary variables $u^\mu$, which are the analogue of the string oscillators (\mt{\a_{-1}^{\mu}\leftrightarrow u^{\mu}}), and define the generating function:
\be
	\varphi^{\sst A}(x,u)=
	\tfrac1{s!}\,\varphi^{\sst A}_{\mu_{1}\cdots\mu_{s}}(x)\,
	u^{\mu_{1}}\cdots u^{\mu_{s}}\,.
\ee
Here, the superscript $\st A$ labels different HS fields, and in the following we use \mt{{\st A}=a} for massless fields and \mt{{\st A}=\a} for massive ones.
In this notation, the most general form of the cubic vertices is
\be\label{gen cub}
	S^{\sst (3)}=\int d^{d}x\,C_{\sst A_{1}A_{2}A_{3}}
	(\partial_{x_{i}},\partial_{u_{i}})\,
	\varphi^{\sst A_{1}}(x_{1},u_{1})\,\varphi^{\sst A_{2}}(x_{2},u_{2})\,
	\varphi^{\sst A_{3}}(x_{3},u_{3})\,
	\Big|_{{}^{x_{i}=x}_{u_{i}=0}}\,,
\ee
where \mt{i=1,2,3}. Different functions $C_{\sst A_{1}A_{2}A_{3}}$ describe 
different vertices, embodying the coupling constants of the theory. Notice that $C_{\sst A_{1}A_{2}A_{3}}$ plays the
same role as the state $\left|V_{3}\right>$ in the BRST (String field theory) approach.
Restricting the attention to the parity invariant interactions, the dependence of
 $C_{\sst A_{1}A_{2}A_{3}}$ on the six vectors $\partial_{x_{i}}$
and $\partial_{u_{i}}$, 
is through the 21 (6+9+6) Lorentz scalars 
\mt{\partial_{x_{i}}\!\cdot\partial_{x_{j}}},
\mt{\partial_{u_{i}}\!\cdot\partial_{x_{j}}}
and \mt{\partial_{u_{i}}\!\cdot\partial_{u_{j}}}.
For instance, a vertex of the form:
\be
	\varphi_{\mu\nu\rho\lambda}\,\partial^\mu\,\partial^\nu\, \varphi^{\rho\sigma}_{\phantom{\rho\sigma}\sigma}\,\partial_\tau\,\varphi^{\tau\lambda}\,,
\ee
is encoded by
\be
	C=(\partial_{u_{1}}\!\cdot\partial_{x_{2}})^2\,(\partial_{u_{1}}\!\cdot\partial_{u_{2}})\,(\partial_{u_{1}}\!\cdot\partial_{u_{3}})\,(\partial_{u_{2}}\!\cdot\partial_{u_{2}})\,(\partial_{u_{3}}\!\cdot\partial_{x_{3}})\,.
\ee
Notice that not all $C_{\sst A_{1}A_{2}A_{3}}$'s  are physically distinguishable
but there exist two kinds of ambiguities. 
The first is due to the triviality of total derivative terms (or integrations by parts).
This ambiguity can be fixed by removing $\partial_{u_{i}}\!\cdot\partial_{x_{i-1}}$
in terms of the other Lorentz scalar operators  as
\be\label {int by part}
	\partial_{u_{i}}\!\cdot\partial_{x_{i-1}}=
	-\,\partial_{u_{i}}\!\cdot\partial_{x_{i+1}}-\partial_{u_{i}}\!\cdot\partial_{x_{i}}
	+\partial_{u_{i}}\!\cdot\partial_{x}\qquad
	[i\simeq i+3]\,.
\ee
The second ambiguity is related to the possibility of performing non-linear field redefinitions which can create  \emph{fictive} interaction terms. 
However, these vertices are all proportional to the linear EOM so that the corresponding ambiguity can be fixed by disregarding the on-shell vanishing vertices.
This amounts to neglect the dependence of the function $C_{\sst A_{1}A_{2}A_{3}}$
on the $\partial_{x_{i}}\!\cdot\partial_{x_{j}}$'s, since the latter can be expressed as
\be
\label{field redef}
	\partial_{x_{i}}\!\cdot\partial_{x_{i-1}}=\tfrac12
	\left(\partial_{x_{i+1}}^{2}\!\!-
	\partial_{x_{i}}^{2}-\partial_{x_{i-1}}^{2}\right)
	+\tfrac12\,\partial_{x}\cdot\!
	\left(\partial_{x_{i}}+\partial_{x_{i-1}}-\partial_{x_{i+1}}\right),
\ee
and, up to EOM, the $\partial_{x_{i}}^{2}$'s can be replaced by the 
$\partial_{u_{i}}\!\cdot\partial_{x_{i}}$'s and the $\partial_{u_{i}}^{2}$'s. For instance, the (Fronsdal's) massless HS EOM reads
\be\label{Fronsdal}
	\left[\partial_{x}^{2}-u\cdot\partial_{x}\,\partial_{u}\cdot\partial_{x}
	+\tfrac12\,(u\cdot\partial_{x})^{2}\,\partial_{u}^{\,2}\right]
	\varphi^{a}(x,u)\approx0\,.
\ee
Taking into account the latter ambiguities,
the vertex function $C_{\sst A_{1}A_{2}A_{3}}$ can only depend 
on 12 (3+3+6) Lorentz scalars:
$\partial_{u_{i}}\!\cdot\partial_{x_{i+1}}$\,,
$\partial_{u_{i}}\!\cdot\partial_{x_{i}}$
and $\partial_{u_{i}}\!\cdot\partial_{u_{j}}$\,.
It is worth noticing that
at the cubic order, there are no non-local vertices 
since the $\partial_{x_{i}}\!\cdot\partial_{x_{j}}$'s have been removed
while the other scalar operators can only enter with 
positive powers (otherwise the tensor contractions do not make sense). 

When a gauge field, say $\varphi^{a_{1}}$, enters the interaction, 
the function $C_{\sst a_{1}A_{2}A_{3}}$ 
is further constrained by the condition \eqref{N2-}.
In this notation, the linear HS gauge transformations, 
\mt{\delta^{\sst (0)}\varphi^{a}_{\mu_{1}\cdots\mu_{s}}=\partial^{}_{(\mu_{1}}\varepsilon^{a}_{\mu_{2}\cdots \mu_{s})}}\,, read
\be\label{free gt}
	\delta^{\sst (0)}\varphi^{a}(x,u)=u\cdot\partial_{x}\,\varepsilon^{a}(x,u)\,.
\ee
Hence the cubic-order gauge consistency condition \eqref{N2-} gives
\be\label{gauge inv C}
	\left[\,C_{a_{1}{\sst A_{2}A_{3}}}(\partial_{u_{i}}\!\cdot\partial_{x_{i+1}}\,,
	\partial_{u_{i}}\!\cdot\partial_{x_{i}}\,,
	\partial_{u_{i}}\!\cdot\partial_{u_{j}})\,,\,
	u_{1}\cdot\partial_{x_{1}}\,\right]\approx 0\,,
\ee
where $\approx$ means equivalence modulo the Fronsdal equation \eqref{Fronsdal}. In order to tackle the above equation,
it is convenient to split the function $C_{a_{1}{\sst A_{2}A_{3}}}$ into two parts:
the one which does not involve any divergence, $\partial_{u_{i}}\!\cdot\partial_{x_{i}}$\,, trace, $\partial_{u_{i}}^{2}$ or auxiliary fields
(we call it the \emph{transverse and traceless} (TT) part),
and the one  which does (DTA part).
Let us notice that the TT part is precisely what survives after eliminating unphysical DOF. Indeed, besides the mass-shell condition, the Fierz system:
\be\label{Fierz}
	{\rm Fierz\ system}:\quad
	(\partial_{x}^{2}-m^{2})\,\varphi^{\sst A}=0\,,
	\quad \partial_{u}\cdot\partial_{x}\,\varphi^{\sst A}=0\,,
	\quad \partial_{u}^{2}\,\varphi^{\sst A}=0\,,
\ee
involves also the transverse  and traceless conditions.\footnote{
When \mt{m=0}, one has to quotient the system by the gauge symmetries
\eqref{free gt} with parameters $\varepsilon^{a}$ satisfying 
the same conditions \eqref{Fierz}.}
Therefore, 
the TT part of the action plays a key role  encoding the on-shell content of the theory.
On the other hand, the part containing divergences, traces or auxiliary fields vanishes after gauge fixing.\footnote{ 
Indeed the light-cone interaction vertices \cite{Metsaev:2005ar} can 
be obtained solely from the TT part by going to the light-cone gauge.}
The next question is whether it is possible to determine the TT part 
of the vertex  without
using any information about the other part.
From a physical point of view, this ought to be possible 
since the physical (on-shell) interactions
cannot depend on the unphysical ones.\footnote{
The light-cone gauge approach is 
consistent in its own without calling for some additional conditions on
its corresponding covariant off-shell description.}
Concerning massive fields, the TT conditions 
already assure the propagation of the correct DOF 
and no further constraints has to be imposed on the TT parts of the cubic interactions.

Let us also comment on the remaining parts  of the cubic interactions involving divergences and traces.
For massless fields, the latter turns out to be completely 
determined by their TT part enforcing gauge invariance \cite{Manvelyan:2010jr,Sagnotti:2010at}.
Similarly, when massive fields are involved, 
after introducing Stueckelberg fields
into the TT part (see \cite{Joung:2012rv} for details),
one may in principle determine  the remaining parts of the action
requiring the consistency with Stueckelberg gauge symmetries.

In the following, we show how to determine the TT parts of 
$C_{\sst A_{1}A_{2}A_{3}}$ from eq.~\eqref{gauge inv C}.
First, after removing all the ambiguities through eqs.~(\ref{int by part}\,,\,\ref{field redef}),
any functional $F$ can be \emph{univocally} written in terms 
of its TT part and the remaining part as
\mt{F=[F]_{\rm\sst TT}+[F]_{\rm\sst DTA}}\,.
Hence, eq.~\eqref{N2-} can be split into two equations:
\be\label{N2--}
	\left[\delta^{\sst (0)}S^{\sst (3)}\right]_{\rm\sst TT} \approx 0\,,
	\qquad \left[\delta^{\sst (0)}S^{\sst (3)}\right]_{\rm\sst DTA} \approx 0\,,
\ee
where, henceforth, $\approx$ means equivalence modulo 
the Fierz system \eqref{Fierz}. 
Second, as the gauge variations of 
divergences, traces or auxiliary fields are proportional to themselves
up to $\partial_{x_{i}}^{2}$-terms:
\mt{\left[\delta^{\sst (0)}[S^{\sst (3)}]_{\rm\sst DTA}\,
\right]_{\rm\sst TT}\approx 0}\,,
the first equation in \eqref{N2--} gives
an independent condition for the TT parts, $[S^{\sst (3)}]_{\rm\sst TT}$\,, of the interactions:
\be\label{N2---}
	\left[\delta^{\sst (0)}S^{\sst (3)}\right]_{\rm\sst TT}
	=\left[\delta^{\sst (0)}\big\{
	\left[S^{\sst (3)}\right]_{\rm\sst TT}
	+\left[S^{\sst (3)}\right]_{\rm\sst DTA}\big\}\right]_{\rm\sst TT}
	\approx
	\left[\delta^{\sst (0)}\left[S^{\sst (3)}\right]_{\rm\sst TT}\right]_{\rm\sst TT}
	\approx 0\,.
\ee
At this point, $[S^{\sst (3)}]_{\rm\sst TT}$ can be expressed
as in eq.~\eqref{gen cub}
through a function $C^{\sst\rm TT}_{\sst A_{1}A_{2}A_{3}}(y_{i},z_{i})$ 
of 6 variables:
\be\label{y and z}
	y_{i}=\partial_{u_{i}}\!\cdot\partial_{x_{i+1}}\,,\qquad
	z_{i}=\partial_{u_{i+1}}\!\!\cdot\partial_{u_{i-1}}\,.
\ee
Then, assuming the first field to be massless, \mt{{\st A_{1}}=a_{1}},
eq.~\eqref{N2---} gives a condition for $C^{\sst\rm TT}_{a_{1}{\sst A_{2}A_{3}}}$
analogous to \eqref{gauge inv C}.
Using the Leibniz rule, we obtain a rather simple differential equation:
\be\label{flat PDE}
	\left[y_{2}\,\partial_{z_{3}}-y_{3}\,\partial_{z_{2}}
	+\tfrac12\,(m_{2}^{\,2}-m_{3}^{\,2})\,\partial_{y_{1}}\right]
	C^{\sst\rm TT}_{a_{1}{\sst A_{2}A_{3}}}=0\,,
\ee
where $m_{i}$ is the mass of the $i$-th field. When two or three massless fields are involved in the interactions, one has respectively two or three differential equations given by the cyclic permutations 
of eq.~\eqref{flat PDE}.

Depending on the cases, the corresponding solutions 
$C^{\sst\rm TT}_{\sst A_{1}A_{2}A_{3}}$ are constrained to depend on
some of the $y_{i}$'s and the $z_{i}$'s only through the \emph{building blocks}:
\be
	g=y_{1}\,z_{1}+y_{2}\,z_{2}+y_{3}\,z_{3}\,,
\ee
\be
\label{flath}
	h_{i}=y_{i+1}\,y_{i-1} +\tfrac12\left[m_{i}^{\,2}-
	(m_{i+1}+m_{i-1})^{2}\right] z_{i}\,.
\ee
As an example, we consider the interactions of three massless HS fields where the consistent cubic interactions are encoded in an arbitrary function:
\be
\label{3flat}
C^{\sst\rm TT}_{a_{1}a_{2}a_{3}}=\mathcal K_{a_{1}a_{2}a_{3}}(y_1,y_2,y_3,g)\,. 
\ee
Leaving aside Chan-Paton factors,
the latter can be expanded as
\be
\label{Kexpan}
\mathcal K=
\sum_{n=0}^{\min\{s_1,s_2,s_3\}}\,\lambda_{n}^{s_1\!-\!s_2\!-\!s_3}\,g^n\,y_1^{s_1-n}\,y_2^{s_2-n}\,y_3^{s_3-n}\,,
\ee
where the \mt{\lambda_{n}^{s_1\!-\!s_2\!-\!s_3}}'s are independent coupling functions that ought to be fixed by the quest for consistency of higher order interactions. From the latter expression it is straightforward to see that the number of consistent couplings is 
\mt{\min\{s_1,s_2,s_3\}+1}\,, while the number of derivatives contained in each vertex is \mt{s_1+s_2+s_3-2\,n}\,. In particular, focussing on the \mt{2\!-\!2\!-\!2} case,
 eq.~\eqref{Kexpan} gives
\be
\cK=\lambda_2^{\sst 2\!-\!2\!-\!2}\,g^2+\lambda_1^{\sst 2\!-\!2\!-\!2}\,g\,y_1\,y_2\,y_3+\lambda_0^{\sst 2\!-\!2\!-\!2}\,y_1^2\,y_2^2\,y_3^2\,,
\ee
that exactly reproduces eq.~\eqref{s2 cubic}.

\section{Ambient-space formalism for HS}
\label{sec: Amb HS}

In order to address the HS interaction problem around an arbitrary
constant-curvature background (\emph{i.e.} (A)dS space),
one can still rely on the Noether procedure introduced in Section \ref{sec: Noether}.
However, in this case the starting point are the HS free theories in (A)dS, where besides massive and massless particles, 
 new types of particles (called partially-massless) appear \cite{Deser:2001us}.\footnote{
 In the case of mixed-symmetry HS fields in AdS,
even the notion of massless-ness changes with respect to the flat-space case \cite{Metsaev:1995re}.}
Moreover, the
cubic interactions built on top of the free theories
would involve (A)dS covariant derivatives whose non-commuting nature makes
 the construction cumbersome. Their commutators give rise to lower-derivative pieces proportional to the cosmological constant, making the vertices inhomogeneous in the number of derivatives. 
The ambient-space formalism proves to be a convenient tool in dealing with free (A)dS HS, and
hence, it represents a natural framework in order to construct (A)dS cubic interactions. Furthermore, recently it has been intensively used in the context of \emph{Mellin amplitude} in the computations of Witten diagrams \cite{Penedones:2010ue,Paulos:2011ie}. 

The ambient-space formalism \cite{Fronsdal:1978vb,Biswas:2002nk} consists in regarding 
the $d$-dimensional (A)dS space as the hyper-surface
 \mt{X^{2}=\epsilon\,L^{2}} embedded into a $(d+1)$-dimensional flat-space.
In our convention the ambient metric
is \mt{\eta=(-,+,\,\ldots,+)}, so that 
AdS (\mt{\epsilon=-1}) is Euclidean while dS (\mt{\epsilon=1}) is Lorentzian. Focussing on the region $\epsilon\,X^{2}>0$\,, there exists an isomorphism between symmetric tensor fields in (A)dS, $\varphi_{\mu_{1}\cdots\mu_{s}}$\,, and those in ambient space, $\Phi_{\sst M_{1}\cdots M_{s}}$\,, satisfying
the \emph{homogeneity} and \emph{tangentiality} (HT) conditions:
\ba
	&{\rm Homogeneity}: \qquad
	&(X\cdot\partial_{X}-U\cdot\partial_{U}+2+\mu)\,\Phi(X,U)=0\,,
	\nn
	&{\rm Tangentiality}: \qquad
	&X\cdot \partial_{U}\,\Phi(X,U)=0\,.	
	\label{HT}
\ea
Here we have used the auxiliary-variable notation for the 
ambient-space fields:
\be
	\Phi(X,U)=\tfrac1{s!}\,
	\Phi_{\sst M_{1}\cdots M_{s}}(X)\,U^{\sst M_{1}}\cdots U^{\sst M_{s}}\,.
\ee
The degree of homogeneity 
$\mu$ parametrizes the (A)dS mass-squared term 
appearing in the (A)dS Lagrangian: 
\be
	m^{2}=\tfrac{(-\epsilon)}{L^{2}}\,\big[\,(\mu-s+2)
	(\mu-s-d+3)-s\,\big]\,,
	\label{mass}
\ee
so that \mt{\mu=0} corresponds to the massless case. 
In the ambient-space formalism, the EOM of both massless and massive HS fields are given by the Fronsdal ones \eqref{Fronsdal} after replacing $(x,u)$ by $(X,U)$\,.   
Let us remind the reader that the concept of \emph{massless-ness} in (A)dS is not related to the
 vanishing of the mass term but rather to 
the appearance of gauge symmetries. In fact, if one postulates the latter to be of the form:
\be\label{amb gauge tr}
	\delta^{\sst (0)}\Phi(X,U)=U\cdot\partial_{X}\,E(X,U)\,,
\ee
 then the compatibility with the HT conditions \eqref{HT} alone
 restricts both the possible values
of $\mu$ and the normal(radial) components of $E$\,. 
In particular, when
\mt{\mu=0,1, \ldots, s-1}, then there exist compatible higher-derivative gauge 
symmetries:
\be
	\delta^{\sst (0)}\,\Phi(X,U)=(U\cdot\partial_X)^{\mu+1}\,\Omega(X,U)
	\qquad
	[\,E=(U\cdot\partial_{X})^{\mu}\,\Omega\,]\,,
\label{pm gt}
\ee 
with the gauge parameters $\Omega$ satisfying
\be
	(X\cdot\partial_X-U\cdot\partial_U-\mu)\,\Omega(X,U)=0\,,
	\qquad X\cdot\partial_U\,\Omega(X,U)=0\,.
\ee
On the other hand, for other values of $\mu$,  no gauge symmetries (in absence of auxiliary fields) are allowed, implying that the corresponding fields 
are massive.
Notice that the massless field, \mt{\mu=0}, is
 the first member of a class of representations where the other members, with \mt{\mu=1,2,\ldots,s-1}, are called partially-massless.
However, partially-massless fields 
describe unitary representations only in dS.

Before closing this section, 
let us discuss the flat limit from the ambient-space viewpoint.
The latter consists first in translating the 
coordinate system as \mt{X^{\sst M}\to X^{M}+L\,N^{\sst M}}\,, where $N$ is a constant vector satisfying $N^{2}=\epsilon$\,, 
and second, in taking  the \mt{L\to \infty} limit.
As a result, the HT conditions \eqref{HT} reduces to
\be
	\left(N\cdot\partial_{X}-\sqrt{-\epsilon}\,M\right) \Phi(X,U)=0\,,
	\qquad
	N\cdot\partial_{U}\,\Phi(X,U)=0\,,
\ee
where the flat mass $M$ is related to the (A)dS \emph{mass} $\mu$ as
\be
\sqrt{-\epsilon}\,M=-\lim_{L\to \infty}\,\frac{\mu}L\,. 
\ee
Notice that in this limit, all (A)dS representations become massless, while, in order to recover massive representations in flat space one should consider the $\mu\rightarrow\infty$ limit.

\section{Ambient-space action}
\label{sec: Amb act}

In the previous section we have shown how to describe HS fields in (A)dS 
making use of the ambient-space language.
In Section \ref{sec: HS cubic} we shall use the latter framework in order to solve the Noether procedure.
For this purpose, one needs to know first of all how to express the (A)dS action 
in terms of ambient-space quantities.
As far as the Lagrangian is concerned, no subtleties arise since, together with the isomorphism between (A)dS and ambient-space fields,  
there is an analogous one between (A)dS-covariant derivatives $\nabla_{\mu}$ and 
 ambient-space ones 
\mt{\partial_{X^{\sst M}}-(X_{\sst M}/X^{2})\,X\cdot\partial_X}.
Hence, any Lagrangian $\cL_{\sst\rm (A)dS}$ written in 
terms of (A)dS intrinsic fields
is in one-to-one correspondence with 
the ambient-space one $\cL_{\sst\rm Amb}$\,.
More precisely, considering a single term in the Lagrangian,
the two descriptions are related by
\be
	\cL{\sst\rm Amb}(\,\Phi,\partial\,\Phi, \partial\,\partial\,\Phi, \ldots\,)=
	\left(\tfrac RL\right)^{\Delta}\,
	\cL_{\sst\rm (A)dS}(\,\varphi,\nabla\varphi,\nabla\nabla\varphi, \ldots\,)\,,
\ee
where $\Delta$ is a constant depending on 
the spins and the $\mu$-values of the fields as well as on the number of derivatives 
entering $\cL_{\rm\sst Amb}$\,.
Regarding the action, the first attempt would be to consider 
\be
	\int d^{d+1}X\,\cL_{\sst\rm Amb}
	=\left(\int_{0}^{\infty} d R \left(\tfrac{R}{L}\right)^{d+\Delta}\right)\times
	\left(\int_{{\rm\sst (A)dS}} d^{d}x\sqrt{-\e\,g}\,\cL_{\rm\sst (A)dS}\right).
\ee
However, the latter contains a diverging radial integral
so that controlling its gauge invariance becomes ambiguous.
A way of solving this problem would be to introduce a cut-off in order to regulate the radial integral, or similarly, a boundary for the ambient space. Then, the presence of the boundary breaks gauge invariance which can be restored only by adding boundary (total-derivative) terms in the action. The latter are the analogue of the Gibbons-Hawking-York boundary term needed in order to amend the Einstein-Hilbert action in manifolds with boundary.

Another equivalent way is suggested by the fact that the ambient space 
can be considered as a tool to rewrite intrinsic $d$-dimensional (A)dS quantities in a manifestly $SO(1,d)$-covariant form. 
In this respect, with the aid of a delta function, one can simply rewrite the (A)dS action in the 
ambient-space language as
\be
	S=\int d^{d}x\,\sqrt{-\epsilon\, g}\,\cL_{\rm\sst (A)dS}=
	\int d^{d+1}X\,\delta\!\left(\sqrt{\epsilon\, X^{2}}-L\right)\,\cL_{\rm\sst Amb}\,.
\ee
As a candidate for the Lagrangian $\cL_{\rm\sst Amb}$\,, one may think to use 
the flat $d$-dimensional one where all the $d$-dimensional quantities are replaced by $(d+1)$-dimensional ones.
However, in general this way does not lead to a consistent (A)dS action. 
The reason is that, because of the delta function, total-derivative terms 
in $\cL_{\sst\rm Amb}$ no longer vanish but contribute as
\be
	\label{intbyparts}
	\delta\!\left(\sqrt{\epsilon\, X^{2}}-L\right) \partial_{X^M}\left(\,\cdots\right)
	=-\,\delta^{\prime}\!\left(\sqrt{\epsilon \,X^{2}}-L\right) 
	\tfrac{\epsilon\,X_M}{\sqrt{\epsilon\,X^{2}}}\,\left(\,\cdots\right)\neq 0\,.
\ee
Thereby, in order to compensate these terms,
the Lagrangian $\cL_{\sst \rm Amb}$ has to be amended 
by additional total-derivative contributions.
It is worth noticing that the latter vertices contain a lower number of derivatives
compared to the initial vertices in $\cL_{\sst \rm Amb}$\,.
Actually, this is the ambient-space analogue of 
what happens in the intrinsic formulation:
the replacement of ordinary derivatives by covariant ones requires 
the inclusion of additional lower-(covariant)derivative vertices
in the Lagrangian.

As previously discussed,
a consistent (A)dS action consists of vertices containing terms with different number
of derivatives sized by proper powers of $L^{-2}$\,: 
\be
	\cL_{\sst\rm Amb}=\cL_{\sst\rm Amb}(L^{-2})\,.
\ee
In the ambient-space formalism, it is convenient to rather 
size such contributions by different derivatives of the delta function:
\be
	\delta^{\sst [n]}\!\left(\sqrt{\epsilon \,X^{2}}-L\right)
	\qquad
	\left[\,\delta^{\sst [n]}(R-L)=\left(\tfrac{1}{R}\,\tfrac{d}{dR}\right)^n\,\delta(R-L)\,\right],
\ee
since the latter naturally appear in the terms \eqref{intbyparts} 
that need to be compensated.
 Indeed, thanks to the following identity:
\be
\label{deltafunc}
	\delta^{\sst [n]}(R-L)\,\,R^{\lambda}
	=\frac{(-2)^{n}\,[(\lambda-1)/2]_{n}}{(L^{2})^{n}}\,\delta(R-L)\,R^{\lambda}\,,
\ee
arbitrary powers of $L^{-2}$ can be always absorbed into derivatives
of the delta function. Therefore, the ambient-space Lagrangian
can be expanded as
\be\label{L series}
	\delta\!\left(\sqrt{\epsilon \,X^{2}}-L\right)\,\cL_{\sst \rm Amb}(L^{-2})
	=\sum_{n\ge0} 
	\delta^{\sst [n]}\!\left(\sqrt{\epsilon \,X^{2}}-L\right)\,\cL_{\sst\rm Amb}^{\sst [n]}\,,
\ee
where the $\cL_{\sst\rm Amb}^{\sst [n]}$'s do not involve any power of $L^{-2}$.
In order to conveniently handle the above series, it is useful to 
express $\delta^{\sst [n]}$ by means of an auxiliary variable 
$\hat{\d}$ as
\be
	\delta^{\sst [n]}(R-L)=\exp\left(\tfrac{\epsilon\,L}{R}\,\tfrac{d}{dR}
	\,\tfrac{d}{d\hat\delta}\right)\,\delta(R-L)\,
	\left(\epsilon\,\tfrac{\hat\delta}L\right)^{n}\,\Big|_{\hat\delta=0}\,.
\ee
For simplicity, in the following  we work with the rule
\mt{\delta^{\sst [n]}(R-L)=\delta(R-L)\,(\epsilon\,\hat{\d}/L)^n}.\footnote{
The $1/L$ in the definition of $\hat\delta$ has been introduced to
provide a well-defined flat limit: the corresponding rule in flat space becomes
\mt{\delta(N\cdot X)\,\partial_{X^{\sst M}}\,(\,\cdots)
	=-\,\delta(N\cdot X)\ \hat\delta\ N_{\sst M}\,(\,\cdots)}\,.}
The advantage of introducing the auxiliary variable $\hat\delta$ lies on the simple rule
in dealing with total derivatives:
\be
	\label{intdeltahat}
	\delta\Big(\sqrt{\epsilon\,X^2}-L\Big)\,\partial_{X^M}\, (\,\cdots)=
	-\,\delta\Big(\sqrt{\epsilon\,X^2}-L\Big)\,\tfrac{\hat\delta}{L}\,X_{\sst M}\,
	(\,\cdots)\,.
\ee
Moreover, it also allows 
one to factorize the delta function in the series \eqref{L series} and rewrite the Lagrangian as a polynomial function in $\hat\delta/L$\,: 
\be
	\cL_{\sst\rm Amb}=\cL_{\sst\rm Amb}(\tfrac{\hat\delta}L)\,.
\ee

\section{Construction of HS cubic interactions in (A)dS}
\label{sec: HS cubic}

In this section we present the solution to the Noether procedure at the cubic level of  
for arbitrary symmetric HS fields in (A)dS. 
The logic is the same as in the flat-space case discussed in Section \ref{sec: flat}, thus in the following
we mainly focus on those points wherein the peculiarities of (A)dS arise.

Apart from the presence of the delta function,
the discussions which led to the most general form of the TT parts of the
cubic vertices still hold. The only subtleties 
are related to the total-derivative terms in \eqref{int by part} and \eqref{field redef}
which no longer vanish. However, as we shall explain below, their contributions
can be reabsorbed into redefinitions of the cubic vertices.
Hence, the most general expression for the TT parts of the cubic vertices reads 
\ba
	\label{cubicact1}
	[S^{\sst {(3)}}]_{\sst\rm TT} \eq
	\int d^{d+1}X\ \delta\Big(\sqrt{\epsilon\,X^2}-L\Big)\,
	C^{\sst\rm TT}_{\sst {A_{1}A_{2}A_{3}}}
	\!\left(\tfrac{\hat\delta}L; Y_{i},Z_{i}\right)\times\nn
	&& \qquad \times\,
	\Phi^{\sst A_{1}}(X_1,U_1)\, \Phi^{\sst A_{2}}(X_2,U_2)\, 
	\Phi^{\sst A_{3}}(X_3,U_3)\, \Big|_{^{ X_i=X}_{ U_i=0}}\,,
\ea
where the $Y_i$'s and the $Z_i$'s are defined analogously to \eqref{y and z}.
Let us remind the reader that, as we have discussed in the previous section,
the inhomogeneity of the vertices in the number of derivatives 
is encoded in the $\hat\delta/L$-dependence of the function 
$C_{\sst A_{1}A_{2}A_{3}}^{\sst\rm TT}$\,. 

Whenever a gauge field joins the interactions,
the cubic vertices are constrained
to satisfy the gauge compatibility condition \eqref{N2---}
associated to that field.
Assuming the first field to be (partially-)massless
(\emph{i.e.} \mt{\mu_{1}\in\{0,1,\ldots,s_{1}-1\}}), one gets 
 \be
\label{gaugeconscond1}
\left[\,C^{\sst\rm TT}_{{a}_{1}{\sst A_{2}A_{3}}}\!\left(\tfrac{\hat\delta}L;Y,Z\right)\,,
\, (U_1\!\cdot\partial_{X_1})^{\mu_1+1}\,\right]\Big|_{U_1=0}\approx 0\,,
\ee
where $\approx$ means again equivalence
modulo the $\partial_{X_{i}}^{\,2}$'s\,, $\partial_{U_{i}}\!\cdot\partial_{X_{i}}$'s
and $\partial_{U_{i}}^{\,2}$'s\,, \emph{i.e.} modulo the ambient-space Fierz system. 
Aside from the higher-derivative nature of the gauge transformations,
the key difference with respect to the flat case is the 
non-triviality of the total-derivative terms arising from the commutations
of \mt{U_{1}\!\cdot\partial_{X_{1}}} with the $Y_{i}$'s and the $Z_{i}$'s.
Let us sketch how these total-derivative terms can be dealt with:
\begin{itemize}
\item
Because of the identity \eqref{intdeltahat}, 
the total-derivatives terms give rise to contributions of order $\hat\delta/L$ and proportional
to the operators \mt{X\cdot\partial_{X_{i}}}
or \mt{X\cdot\partial_{U_{i}}}\,. 
\item 
Appearing right after the delta function,
the latter can be replaced by
\mt{X_{i}\cdot\partial_{X_{i}}} and \mt{X_{i}\cdot\partial_{U_{i}}} respectively.
\item
Pushing these operators to the right and making them act on the fields, one can use the HT conditions \eqref{HT}
to replace \mt{X_{i}\!\cdot\partial_{X_{i}}}  by the corresponding homogeneity degrees
and \mt{X_{i}\!\cdot\partial_{U_{i}}}  by zero. 
\end{itemize}
All in all,
one can recast the condition \eqref{gaugeconscond1} into a higher-order partial differential equation of the form:
\be
\label{pmdiff}
\prod_{n=0}^{\mu_1}\Big[\,Y_2\,\partial_{Z_3}-Y_3\,\partial_{Z_2}
+\tfrac{\hat\delta}{L}\left(Y_2\,\partial_{Y_2}-Y_3\,\partial_{Y_3}
-\tfrac{\mu_1+\mu_2-\m_3-2\,n}2
\right)\partial_{Y_{1}}\Big]\, C^{\sst\rm TT}_{a_{1}\sst A_{2}A_{3}}\!\left(\tfrac{\hat\delta}L;Y,Z\right)=0\,.
\ee
It is worth noticing that the \emph{masses} of the other two fields,
$\mu_{2}$ and $\mu_{3}$\,,
enter the equation as \emph{effective masses},
\mt{\mu_{2}-2\,Y_2\,\partial_{Y_2}} and \mt{\mu_{3}-2\,Y_3\,\partial_{Y_3}}\,,
dressed by number operators. Therefore, even in the massless case
(\mt{\mu_{2}=\mu_{3}=0}) a mass-like term survives. 
Again, depending on the number of (partially-)massless fields 
involved in the interactions,
one can have a system of (up to three) differential equations given by the cyclic permutations 
of eq.~\eqref{pmdiff}.

\section{Solutions of HS cubic interactions in (A)dS}
\label{sec: sol}

In this section we discuss the polynomial solutions of the system of PDEs given by 
eq.~\eqref{pmdiff} and possible cyclic permutations thereof. Indeed, since the generating function $\Phi(X,U)$ is a formal series in $U^{\sst M}$\,, the latter are the only relevant ones.
Our discussion mainly focuses on the interactions involving three massless fields which are of capital importance due to their connections to VE. 

\subsection{Three massless case}

In the three massless case (\mt{\mu_{i}=0}\,, \mt{i=1,2,3}), 
one has a system of three second order PDEs of the form:
\be\label{3massAdS}
\Big[\,Y_{{i+1}}\,\partial_{Z_{{i-1}}}\!-Y_{i-1}\,\partial_{Z_{{i+1}}}
+\tfrac{\hat\delta}{L}\left(Y_{{i+1}}\,\partial_{Y_{{i+1}}}-Y_{{i-1}}\,\partial_{Y_{{i-1}}}
\right)\partial_{Y_{i}}\Big]\, C^{\sst\rm TT}_{a_{1} a_{2}a_{3}}\!\left(\tfrac{\hat\delta}L;Y,Z\right)=0\,,
\ee
where \mt{[i\simeq i+3]}\,. The latter can be solved via standard techniques (the Laplace transform
and the method of characteristics), and its solutions are given by
\ba
\label{solansatz}
C^{\sst\rm TT}_{ a_{1}a_{2}a_{3}}\eq
\exp\left\{-\tfrac{\hat\delta}{L}\,\left[Z_1\,\partial_{Y_2}\,\partial_{Y_3}+Z_1\,Z_2\,\partial_{Y_3}\,\partial_{G}+\mbox{cyclic}+Z_1\,Z_2\,Z_3\,\partial_G^2\right]\right\}
\nn
&& \ \times\,\cK_{a_{1}a_{2}a_{3}}(Y_1,Y_2,Y_3,G)\,,
\ea
where $\cK_{a_{1}a_{2}a_{3}}$ is an arbitrary polynomial function of the $Y_i$'s and
 \mt{G=Y_1\,Z_1+Y_2\,Z_2+Y_3\,Z_3}\,. Notice that in the flat limit, one 
 recovers the coupling \eqref{3flat}. The exponential function provides the correct lower-derivative tails needed for the consistency of the corresponding (A)dS interactions. For instance, considering 
the lowest-derivative $4\!-\!4\!-\!4$ interaction, 
\mt{\cK=\lambda_{\sst 4}^{\sst 4\!-\!4\!-\!4}\,G^{4}}\,, one gets
\ba\label{444}
	C^{\sst\rm TT}\eq \lambda_{\sst 4}^{\sst 4\!-\!4\!-\!4}\left[
	G^{4}-12\,\tfrac{\hat\delta}L\,Z_{1}\,Z_{2}\,Z_{3}\,G^{2}
	+12 \left(\tfrac{\hat\delta}L\right)^{\!2}\,Z_{1}^{2}\,Z_{2}^{2}\,Z_{3}^{2}
	\right]\nn
	\eq \lambda_{\sst 4}^{\sst 4\!-\!4\!-\!4}\left[G^{4}+\tfrac{12\,\epsilon\,(d+3)}{L^{2}}\,Z_{1}\,Z_{2}\,Z_{3}\,G^{2}
	+\tfrac{12\,(d+3)(d+5)}{L^{4}}\,Z_{1}^{2}\,Z_{2}^{2}\,Z_{3}^{2}\right]\,,
\ea
where, in the second line we have used the identity \eqref{deltafunc} in order to replace the powers of $\hat\delta/L$
by those of $L^{-2}$. 

It is worth mentioning another way of presenting the solution \eqref{solansatz}.
It consists in encoding all the $\hat{\delta}$ contributions into total derivatives 
as 
\be
	C^{\sst \rm TT}_{a_{1}a_{2}a_{3}}=
	\cK_{a_{1}a_{2}a_{3}}\big(\tilde{Y}_{1},\tilde{Y}_{2},\tilde{Y}_{2},\tilde{G}\big)\,,
	\label{amb K}
\ee
where the $\tilde{Y}_i$'s and $\tilde{G}$ are the (A)dS
deformations:
\ba	\label{amb G}
	&& \tilde{Y}_{i} = Y_{i}+\alpha_{i}\,\partial_{U_{i}}\cdot\partial_{X}\,, \
	\qquad\tilde{G} = \sum_{i=1}^{3}
	(Y_{i} + \beta_{i}\,\partial_{U_{i}}\cdot\partial_{X})\,Z_{i}\,,\nn
	&&
	 (\alpha_{1},\alpha_{2},\alpha_{3}\,;\,\b_{1},\b_{2},\b_{3})
	 =(\alpha,-\tfrac{1}{\alpha+1},-\tfrac{\alpha+1}{\alpha}\,;\,
	 \beta,-\,\tfrac{\beta+1}{\alpha+1},-\,\tfrac{\alpha-\beta}{\alpha})\,,
\ea
of the flat-space building blocks $Y_i$'s and $G$\,.
The equivalence between the two representations  
\eqref{solansatz} and \eqref{amb K} 
of the cubic interactions can be shown carrying out the integration by parts of 
all the total-derivative terms present in \eqref{amb K}. 
Notice that the freedom of $\alpha$ and $\beta$ reflects
a redundancy in expressing the building blocks 
in terms of total derivatives.
Finally, let us conclude the discussion on the interactions of massless HS fields 
providing the example \eqref{444} in terms of ambient-space tensor contractions:
\ba
	&&[S^{\sst (3)}]_{\sst\rm TT}=\lambda_{\sst 4}^{\sst 4\!-\!4\!-\!4}\,\int d^{d+1}X\ \delta\Big(\sqrt{\epsilon\,X^2}-L\Big)\,\nn
	&&\times\Big[\Phi_{\sst MNPQ}\,\partial^{\sst M}\,\partial^{\sst N}\,\partial^{\sst P}\,\partial^{\sst Q}\,\Phi_{\sst RSTV}\,\Phi^{\sst RSTV}+8\,\Phi_{\sst MNPQ}\,\partial^{\sst M}\,\partial^{\sst N}\,\partial^{\sst P}\,\Phi_{\sst RSTV}\,\partial^{\sst V}\,\Phi^{\sst RSTQ}\nn
	&&\quad\ +\,6\,\Phi_{\sst MNTV}\,\partial^{\sst M}\,\partial^{\sst N}\,\Phi_{\sst PQRS}\,\partial^{\sst R}\,\partial^{\sst S}\,\Phi^{\sst PQTV}+12\,\partial_{\sst S}\,\Phi_{\sst MNR}^{\phantom{\sst MNR}\sst T}\,\partial^{\sst M}\,\partial^{\sst N}\,\Phi_{\sst PQRT}\,\partial^{\sst R}\,\Phi^{\sst PQRS}\nn
&&\quad\ 	+\,\tfrac{12\,\epsilon\,(d+3)}{L^{2}}\,\Big(\Phi_{\sst MNS}^{\phantom{\sst MNS}\sst T}\,\partial^{\sst M}\,\partial^{\sst N}\,\Phi_{\sst PQRT}\,\Phi^{\sst PQRS}+2\,\Phi_{\sst MRS}^{\phantom{\sst MRS}\sst T}\,\partial^{\sst M}\,\Phi_{\sst NPQT}\,\partial^{\sst N}\,\Phi^{\sst PQRS}\Big)\nn
	&&\quad\ +\,\tfrac{4\,(d+3)(d+5)}{L^{4}}\,\Phi_{\sst MNPQ}\,\Phi^{\sst MNRS}\,\Phi^{\sst PQ}_{\phantom{\sst PQ}\sst RS}\Big]\,.	
\ea

\subsection{General cases}

The interactions of three massless fields represent a subclass of 
the interactions one can envisage depending on the values of the $\mu_{i}$'s\,.
Let us notice that  in the general cases the solutions
are given by intersections of the solution spaces of the PDE \eqref{pmdiff}
and its cyclic permutations.
Therefore, we start our discussion from the solutions of one PDE,
for which it is instructive to first analyze 
the corresponding equation in flat space \eqref{flat PDE}.
The latter exhibits a singular point in correspondence of the value \mt{m_2=m_3}\,. 
Indeed, aside from this value, 
a rescaling of \mt{m_2^2-m_3^2}  
is tantamount to a rescaling of $y_1$\,.
Therefore, any polynomial solution with \mt{m_2\neq m_3} can be smoothly deformed to a solution with \mt{m_2=m_3}\,, while the opposite is not true.
Consequently, the solution space with \mt{m_2=m_3}
is always bigger than (or equal to) the one with \mt{m_2\neq m_3}\,.
Indeed, an explicit analysis shows that  
the \mt{m_{2}\neq m_{3}} solutions \mt{\cK(y_{2},y_{3},h_{2},h_{3},z_{1})}
can be always expressed in terms of the \mt{m_{2}=m_{3}} solutions
\mt{\cK(y_{1},y_{2},y_{3},g,z_{1})},
while the opposite is not true.

The latter phenomenon has a richer counterpart in (A)dS,
where  the role of \mt{m_{2}^{2}-m_{3}^{2}}  in eq.~\eqref{flat PDE} is played by
the combinations:
 \be\label{mu value}
 	\mu_1+\mu_2-\m_3-2\,(Y_2\,\partial_{Y_2}-Y_3\,\partial_{Y_3}+n)
	\qquad [n=0,\ldots,\mu_{1}]\,,
\ee
in  eq.~\eqref{pmdiff}. Indeed,
because of the number operator \mt{Y_2\,\partial_{Y_2}-Y_3\,\partial_{Y_3}},
eq.~\eqref{mu value} may have several vanishing points for
 \mt{\mu_1+\mu_2-\m_3\in 2\,\mathbb{Z}}\,. 
More precisely, in correspondence of the latter values, one can consider an ansatz of the form: 
\be
\label{Y3ans}
C^{\sst\rm TT}_{a_{1}\sst A_{2}A_{3}}\!\left(\tfrac{\hat\delta}L;Y,Z\right)=Y_{2,3}^{|M|}\,\bar{C}^{\sst\rm TT}_{a_{1}\sst A_{2}A_{3}}\!\left(\tfrac{\hat\delta}L;Y,Z\right)\qquad
\left[M=\tfrac{\mu_1+\mu_2-\m_3-2\,n}2\right],
\ee
where we use $Y_{2}$ for \mt{M>0}\, and
$Y_{3}$ for \mt{M<0}\,. 
Plugging this ansatz into the original equation \eqref{pmdiff},
one ends up with an analogous equation for $\bar{C}^{\sst\rm TT}_{a_{1}\sst A_{2}A_{3}}$\,, 
whose $n$-th factor coincides with the operator appearing in 
the massless case \eqref{3massAdS}.
Therefore, the solutions of the massless equation provide
solutions of the original equation through the ansatz \eqref{Y3ans}.
Notice that, when $\mu_1+\mu_2-\m_3\notin 2\,\mathbb{Z}$\,, 
the aforementioned solutions are no longer available since they become non-polynomial.
In all the cases which can not be covered by the ansatz \eqref{Y3ans}, the solutions can be expressed as arbitrary functions of the building blocks: 
\be
\tilde H_i=\partial_{U_{i-1}}\!\cdot\partial_{X_{i+1}}\,\partial_{U_{i+1}}\!\cdot\partial_{X_{i-1}}\!-\partial_{X_{i+1}}\!\cdot\partial_{X_{i-1}}\,Z_i\,,
\ee 
which are the (A)dS deformations of the flat-space building blocks $h_{i}$ \eqref{flath}. 
It is worth noticing that this pattern is similar to what happens
in flat space where
the $h_{i}$-type solutions exist independently on the mass values,
while the massless-type ones (involving $g$) only appear 
for particular values of the $m_{i}$'s.

Moving to the cases in which more than one equation is involved,
one has to consider intersections of the corresponding solution spaces.
Since in flat space the only \emph{enhancement point}
arises for \mt{m_{i}=m_{i+1}}\,,
one is led to five different cases:
(1) three massless (\mt{m_{1}=m_{2}=m_{3}=0}),
(2) two massless and one massive (\mt{m_{1}=m_{2}=0\,, m_{3}\neq 0}), 
(3) one massless and two massive with different masses 
(\mt{m_{1}=0\,, m_{2}\neq m_{3}}),
(4) one massless and two massive with equal masses (\mt{m_{1}=0\,, m_{2}=m_{3}}),
(5) three massive.
On the other hand, due to the presence of a richer pattern
of enhancement points (\mt{\mu_i+\mu_{i+1}-\m_{i-1}\in 2\,\mathbb{Z}}), 
more combinations appear in (A)dS. 
The analysis of the above cases goes beyond the scope of the present letter, 
and we refer to the forthcoming paper \cite{Joung:2012} for the detailed discussion.

\acknowledgments{
We are grateful to
D. Francia, K. Mkrtchyan and A. Sagnotti
for helpful discussions.
The present research was supported in part by Scuola Normale Superiore, by INFN (I.S. TV12) and by the MIUR-PRIN contract 2009-KHZKRX.}

\providecommand{\href}[2]{#2}\begingroup\raggedright\endgroup


\begin{thebibliography}{10}

\bibitem{Vasiliev:1988sa}
M.~A. Vasiliev, {\it {Consistent equations for interacting massless fields of
  all spins in the first order in curvatures}},  {\em Annals Phys.} {\bf 190}
  (1989) 59--106.

\bibitem{Vasiliev:2003ev}
M.~Vasiliev, {\it {Nonlinear equations for symmetric massless higher spin
  fields in (A)dS(d)}},  {\em Phys.Lett.} {\bf B567} (2003) 139--151,
  [\href{http://xxx.lanl.gov/abs/hep-th/0304049}{{\tt hep-th/0304049}}].

\bibitem{Bekaert:2010hw}
X.~Bekaert, N.~Boulanger, and P.~Sundell, {\it {How higher-spin gravity
  surpasses the spin two barrier: no-go theorems versus yes-go examples}},
  \href{http://xxx.lanl.gov/abs/1007.0435}{{\tt arXiv:1007.0435}}.

\bibitem{Sagnotti:2011qp}
A.~Sagnotti, {\it {Notes on Strings and Higher Spins}},
  \href{http://xxx.lanl.gov/abs/1112.4285}{{\tt arXiv:1112.4285}}.

\bibitem{Taronna:2010qq}
M.~Taronna, {\it {Higher Spins and String Interactions}},
  \href{http://xxx.lanl.gov/abs/1005.3061}{{\tt arXiv:1005.3061}}.

\bibitem{Sagnotti:2010at}
A.~Sagnotti and M.~Taronna, {\it {String Lessons for Higher-Spin
  Interactions}},  {\em Nucl. Phys.} {\bf B842} (2011) 299--361,
  [\href{http://xxx.lanl.gov/abs/1006.5242}{{\tt arXiv:1006.5242}}].

\bibitem{Sezgin:2002ru}
E.~Sezgin and P.~Sundell, {\it {Analysis of higher spin field equations in
  four-dimensions}},  {\em JHEP} {\bf 0207} (2002) 055,
  [\href{http://xxx.lanl.gov/abs/hep-th/0205132}{{\tt hep-th/0205132}}].

\bibitem{Sezgin:2003pt}
E.~Sezgin and P.~Sundell, {\it {Holography in 4D (super) higher spin theories
  and a test via cubic scalar couplings}},  {\em JHEP} {\bf 0507} (2005) 044,
  [\href{http://xxx.lanl.gov/abs/hep-th/0305040}{{\tt hep-th/0305040}}].

\bibitem{Boulanger:2008tg}
N.~Boulanger, S.~Leclercq, and P.~Sundell, {\it {On The Uniqueness of Minimal
  Coupling in Higher-Spin Gauge Theory}},  {\em JHEP} {\bf 0808} (2008) 056,
  [\href{http://xxx.lanl.gov/abs/0805.2764}{{\tt arXiv:0805.2764}}].

\bibitem{Joung:2011ww}
E.~Joung and M.~Taronna, {\it {Cubic interactions of massless higher spins in
  (A)dS: metric-like approach}},  {\em Nucl.Phys.} {\bf B861} (2012) 145--174,
  [\href{http://xxx.lanl.gov/abs/1110.5918}{{\tt arXiv:1110.5918}}].

\bibitem{Joung:2012rv}
E.~Joung, L.~Lopez, and M.~Taronna, {\it {On the cubic interactions of massive
  and partially-massless higher spins in (A)dS}},  {\em JHEP} {\bf 1207} (2012)
  041, [\href{http://xxx.lanl.gov/abs/1203.6578}{{\tt arXiv:1203.6578}}].

\bibitem{Fradkin:1986qy}
E.~Fradkin and M.~A. Vasiliev, {\it {Cubic Interaction in Extended Theories of
  Massless Higher Spin Fields}},  {\em Nucl.\ Phys.} {\bf B291} (1987) 141.

\bibitem{Vasilev:2011xf}
M.~Vasiliev, {\it {Cubic Vertices for Symmetric Higher-Spin Gauge Fields in
  $(A)dS_d$}},  {\em Nucl.Phys.} {\bf B862} (2012) 341--408,
  [\href{http://xxx.lanl.gov/abs/1108.5921}{{\tt arXiv:1108.5921}}].

\bibitem{Buchbinder:2006eq}
I.~Buchbinder, A.~Fotopoulos, A.~C. Petkou, and M.~Tsulaia, {\it {Constructing
  the cubic interaction vertex of higher spin gauge fields}},  {\em Phys.Rev.}
  {\bf D74} (2006) 105018, [\href{http://xxx.lanl.gov/abs/hep-th/0609082}{{\tt
  hep-th/0609082}}].

\bibitem{Zinoviev:2008ck}
Y.~M. Zinoviev, {\it {On spin 3 interacting with gravity}},  {\em Class. Quant.
  Grav.} {\bf 26} (2009) 035022, [\href{http://xxx.lanl.gov/abs/0805.2226}{{\tt
  arXiv:0805.2226}}].

\bibitem{Zinoviev:2010cr}
Y.~Zinoviev, {\it {Spin 3 cubic vertices in a frame-like formalism}},  {\em
  JHEP} {\bf 1008} (2010) 084, [\href{http://xxx.lanl.gov/abs/1007.0158}{{\tt
  arXiv:1007.0158}}].

\bibitem{Fotopoulos:2010nj}
A.~Fotopoulos and M.~Tsulaia, {\it {Current Exchanges for Reducible Higher Spin
  Modes on AdS}},  \href{http://xxx.lanl.gov/abs/1007.0747}{{\tt
  arXiv:1007.0747}}.

\bibitem{Bekaert:2010hk}
X.~Bekaert and E.~Meunier, {\it {Higher spin interactions with scalar matter on
  constant curvature spacetimes: conserved current and cubic coupling
  generating functions}},  {\em JHEP} {\bf 11} (2010) 116,
  [\href{http://xxx.lanl.gov/abs/1007.4384}{{\tt arXiv:1007.4384}}].

\bibitem{Polyakov:2012qj}
D.~Polyakov, {\it {Higher Spin Holography and AdS String Sigma-Model}},
  \href{http://xxx.lanl.gov/abs/1207.4751}{{\tt arXiv:1207.4751}}.

\bibitem{Giombi:2011rz}
S.~Giombi, S.~Prakash, and X.~Yin, {\it {A Note on CFT Correlators in Three
  Dimensions}},  \href{http://xxx.lanl.gov/abs/1104.4317}{{\tt
  arXiv:1104.4317}}.

\bibitem{Costa:2011mg}
M.~S. Costa, J.~Penedones, D.~Poland, and S.~Rychkov, {\it {Spinning Conformal
  Correlators}},  {\em JHEP} {\bf 1111} (2011) 071,
  [\href{http://xxx.lanl.gov/abs/1107.3554}{{\tt arXiv:1107.3554}}].

\bibitem{Costa:2011dw}
M.~S. Costa, J.~Penedones, D.~Poland, and S.~Rychkov, {\it {Spinning Conformal
  Blocks}},  {\em JHEP} {\bf 1111} (2011) 154,
  [\href{http://xxx.lanl.gov/abs/1109.6321}{{\tt arXiv:1109.6321}}].

\bibitem{SimmonsDuffin:2012uy}
D.~Simmons-Duffin, {\it {Projectors, Shadows, and Conformal Blocks}},
  \href{http://xxx.lanl.gov/abs/1204.3894}{{\tt arXiv:1204.3894}}.

\bibitem{Stanev:2012nq}
Y.~S. Stanev, {\it {Correlation Functions of Conserved Currents in Four
  Dimensional Conformal Field Theory}},
  \href{http://xxx.lanl.gov/abs/1206.5639}{{\tt arXiv:1206.5639}}.

\bibitem{Zhiboedov:2012bm}
A.~Zhiboedov, {\it {A note on three-point functions of conserved currents}},
  \href{http://xxx.lanl.gov/abs/1206.6370}{{\tt arXiv:1206.6370}}.

\bibitem{Berends:1984rq}
F.~A. Berends, G.~Burgers, and H.~van Dam, {\it {On the theoretical problems in
  constructing interactions involving higher spin massless particles}},  {\em
  Nucl.\ Phys.} {\bf B260} (1985) 295.

\bibitem{Taronna:2011kt}
M.~Taronna, {\it {Higher-Spin Interactions: four-point functions and beyond}},
  {\em JHEP} {\bf 1204} (2012) 029,
  [\href{http://xxx.lanl.gov/abs/1107.5843}{{\tt arXiv:1107.5843}}].

\bibitem{Dempster:2012vw}
P.~Dempster and M.~Tsulaia, {\it {On the Structure of Quartic Vertices for
  Massless Higher Spin Fields on Minkowski Background}},
  \href{http://xxx.lanl.gov/abs/1203.5597}{{\tt arXiv:1203.5597}}.

\bibitem{Fotopoulos:2010ay}
A.~Fotopoulos and M.~Tsulaia, {\it {On the Tensionless Limit of String theory,
  Off - Shell Higher Spin Interaction Vertices and BCFW Recursion Relations}},
  \href{http://xxx.lanl.gov/abs/1009.0727}{{\tt arXiv:1009.0727}}.

\bibitem{Buchbinder:2012iz}
I.~Buchbinder, T.~Snegirev, and Y.~Zinoviev, {\it {Cubic interaction vertex of
  higher-spin fields with external electromagnetic field}},
  \href{http://xxx.lanl.gov/abs/1204.2341}{{\tt arXiv:1204.2341}}.

\bibitem{Metsaev:2012uy}
R.~Metsaev, {\it {BRST-BV approach to cubic interaction vertices for massive
  and massless higher-spin fields}},
  \href{http://xxx.lanl.gov/abs/1205.3131}{{\tt arXiv:1205.3131}}.

\bibitem{Henneaux:2012wg}
M.~Henneaux, G.~Lucena~Gomez, and R.~Rahman, {\it {Higher-Spin Fermionic Gauge
  Fields and Their Electromagnetic Coupling}},
  \href{http://xxx.lanl.gov/abs/1206.1048}{{\tt arXiv:1206.1048}}.

\bibitem{Metsaev:2005ar}
R.~Metsaev, {\it {Cubic interaction vertices of massive and massless higher
  spin fields}},  {\em Nucl.Phys.} {\bf B759} (2006) 147--201,
  [\href{http://xxx.lanl.gov/abs/hep-th/0512342}{{\tt hep-th/0512342}}].

\bibitem{Manvelyan:2010jr}
R.~Manvelyan, K.~Mkrtchyan, and W.~Ruehl, {\it {General trilinear interaction
  for arbitrary even higher spin gauge fields}},  {\em Nucl.\ Phys.} {\bf B836}
  (2010) 204--221, [\href{http://xxx.lanl.gov/abs/1003.2877}{{\tt
  arXiv:1003.2877}}].

\bibitem{Deser:2001us}
S.~Deser and A.~Waldron, {\it {Partial masslessness of higher spins in (A)dS}},
   {\em Nucl.Phys.} {\bf B607} (2001) 577--604,
  [\href{http://xxx.lanl.gov/abs/hep-th/0103198}{{\tt hep-th/0103198}}].

\bibitem{Metsaev:1995re}
R.~Metsaev, {\it {Massless mixed symmetry bosonic free fields in d-dimensional
  anti-de Sitter space-time}},  {\em Phys.Lett.} {\bf B354} (1995) 78--84.

\bibitem{Penedones:2010ue}
J.~Penedones, {\it {Writing CFT correlation functions as AdS scattering
  amplitudes}},  {\em JHEP} {\bf 1103} (2011) 025,
  [\href{http://xxx.lanl.gov/abs/1011.1485}{{\tt arXiv:1011.1485}}].

\bibitem{Paulos:2011ie}
M.~F. Paulos, {\it {Towards Feynman rules for Mellin amplitudes}},  {\em JHEP}
  {\bf 1110} (2011) 074, [\href{http://xxx.lanl.gov/abs/1107.1504}{{\tt
  arXiv:1107.1504}}].

\bibitem{Fronsdal:1978vb}
C.~Fronsdal, {\it {Singletons and Massless, Integral Spin Fields on de Sitter
  Space (Elementary Particles in a Curved Space. 7.)}},  {\em Phys.\ Rev.} {\bf
  D20} (1979) 848--856.

\bibitem{Biswas:2002nk}
T.~Biswas and W.~Siegel, {\it {Radial dimensional reduction: Anti-de Sitter
  theories from flat}},  {\em JHEP} {\bf 0207} (2002) 005,
  [\href{http://xxx.lanl.gov/abs/hep-th/0203115}{{\tt hep-th/0203115}}].

\bibitem{Joung:2012}
E.~Joung, L.~Lopez, and M.~Taronna. To appear.

\end{thebibliography}
\end{document}